\documentclass[preprintnumbers,amsmath,amssymbm,prd]{revtex4}
\usepackage{epsfig}
\usepackage{graphicx}
\usepackage{amssymb}

\begin{document}
\title{Spontaneous scalarization of charged Reissner-Nordstr\"om black holes: Analytic treatment
along the existence line}
\author{Shahar Hod}
\affiliation{The Ruppin Academic Center, Emeq Hefer 40250, Israel}
\affiliation{ }
\affiliation{The Hadassah Institute, Jerusalem 91010, Israel}
\date{\today}

\begin{abstract}
\ \ \ It has recently been demonstrated that charged black holes can
support spatially regular matter configurations made of massless
scalar fields which are non-minimally coupled to the electromagnetic
field of the charged spacetime. Intriguingly, using numerical
techniques, it has been revealed that the resonant spectra of the
composed charged-black-hole-nonminimally-coupled-scalar-field
configurations are characterized by charge-dependent discrete
scalarization bands $\alpha\in\{[\alpha^{-}_{n}({\bar
Q}),\alpha^{+}_{n}({\bar Q}]\}_{n=0}^{n=\infty}$, where $\alpha$ is
the dimensionless coupling constant of the theory and ${\bar
Q}\equiv Q/M$ is the dimensionless charge-to-mass ratio of the
central supporting black hole. In the present paper we use {\it
analytical} techniques in order to study the physical and
mathematical properties of the spatially regular non-minimally
coupled scalar field configurations (linearized scalar clouds) which
are supported by the central charged Reissner-Nordstr\"om black
holes. In particular, we derive a remarkably compact formula for the
discrete resonant spectrum $\{\alpha^-_n({\bar
Q})\}_{n=0}^{n=\infty}$ which characterizes the composed
black-hole-linearized-field configurations along the {\it
existence-line} of the system, the critical line which separates
bare Reissner-Nordstr\"om black holes from hairy scalarized
black-hole configurations. The analytical results are confirmed by
direct numerical computations.
\end{abstract}
\bigskip
\maketitle

\section{Introduction}

Spherically symmetric charged black holes cannot support
asymptotically flat static or stationary scalar field configurations
with minimal coupling to gravity. This physically important
statement is the conclusion of the mathematically elegant no-hair
theorems presented in \cite{Bek1,Her1,Hodstationary}. The regime of validity
of this intriguing black-hole property has also been extended to the
physically interesting case of scalar fields which are characterized by a
non-minimal coupling to the Ricci scalar of the curved spacetime
\cite{BekMay,Hod1}.

In a very interesting work \cite{Herch}, it has recently been
revealed that asymptotically flat charged black-hole spacetimes {\it
can} support hairy configurations which are made of massless scalar
fields with non-trivial (non-minimal) couplings to the
electromagnetic field tensor of the corresponding central charged
black holes. In particular, it has been shown in \cite{Herch,Herr}
that charged Reissner-Nordstr\"om solutions of the Einstein
equations in field theories whose actions contain a coupling term of
the form $f(\phi)F_{\mu\nu}F^{\mu\nu}$ [here $\phi$ is the scalar
field and $F^{\mu\nu}$ is the Maxwell tensor, see Eq. (\ref{Eq3})
below] may become unstable to perturbations of the non-minimally
coupled scalar fields. The instability of the bare ($\phi\equiv0$)
Reissner-Nordstr\"om black holes signals the formation of hairy
(with $\phi\neq0$) scalar configurations around the central
supporting black holes \cite{Herch,Herr}, a phenomenon which is
known by the name black-hole spontaneous scalarization.

The non-trivial coupling function $f(\phi)$ between the supported
scalar field configurations and the electromagnetic tensor of the
charged black-hole spacetime should be of a mathematical form that
allows, in the limiting trivial $\phi\equiv0$ case, the existence of
the familiar charged Reissner-Nordstr\"om black-hole solutions of
the Einstein field equations \cite{Herch,Herr}. As proved in
\cite{Herch,Herr}, this physically motivated requirement leads to
the universal functional behavior $f(\phi)=1-\alpha\phi^2+O(\phi^4)$
of the coupling function in the linearized regime of the supported
scalar configurations (the term scalar `clouds' is usually used in
the physics literature \cite{Hodlit,Herlit} in order to describe the
external hairy configurations in the linearized weak-field limit).
The physical parameter $\alpha>0$ is the dimensionless coupling
constant of the composed black-hole-scalar-field system.

Intriguingly, using numerical techniques, it has been found in
\cite{Herch,Herr} that the composed black-hole-scalar-field hairy
configurations are characterized by a charge-dependent discrete set
$\alpha\in\{[\alpha^{-}_{n}(Q/M),\alpha^{+}_{n}(Q/M)]\}_{n=0}^{n=\infty}$
of dimensionless scalarization bands \cite{NoteMQ}. In particular,
the discrete eigenvalues $\{\alpha^-_n(Q/M)\}_{n=0}^{n=\infty}$ of
the physical parameter $\alpha$ correspond to charged
Reissner-Nordstr\"om black holes that support {\it linearized}
spatially regular configurations of the massless scalar fields.

The important numerical results presented in \cite{Herch,Herr} can also be
analyzed from a different physical point of view: they reveal the interesting
fact that, for a given value of the dimensionless coupling constant
$\alpha$ which characterizes the field theory, spatially regular
hairy black-hole-scalar-field configurations only exist above some
critical threshold value ${\bar Q}_{\text{min}}(\alpha)$ of the
dimensionless charge-to-mass ratio which characterizes the
supporting Reissner-Nordstr\"om black holes (here ${\bar Q}\equiv
Q/M$ is the charge-to-mass ratio of the central black hole).

The full set of black-hole-linearized-scalar-field configurations
with ${\bar Q}={\bar Q}_{\text{min}}(\alpha)$ make up the {\it
existence line} \cite{Herch,Herr} of the composed physical system in
the parameter plane $({\bar Q},\alpha)$. As discussed in
\cite{Herch,Herr}, this unique critical line, which separates bare
Reissner-Nordstr\"om black holes from hairy (scalarized) black-hole
configurations, is universal in the sense that the various coupling
functions $\{f(\phi)\}$ studied numerically in \cite{Herch,Herr} are
identical in the linearized weak-field regime.

The main goal of the present paper is to study {\it analytically}
the physical and mathematical properties of the cloudy
charged-black-hole-linearized-scalar-field configurations. In
particular, below we shall derive a remarkably compact analytical
formula [see Eq. (\ref{Eq17}) below] for the existence line ${\bar
Q}={\bar Q}_{\text{min}}(\alpha)$ \cite{Notenman} which
characterizes the composed
Reissner-Nordstr\"om-black-hole-nonminimally-coupled-scalar-field
system.

\section{Description of the system}

We shall analyze the physical and mathematical properties of the
composed
Reissner-Nordstr\"om-black-hole-linearized-massless-scalar-field
system which is characterized by a non-trivial coupling of the
scalar field to the electromagnetic field of the charged spacetime.
The curved black-hole spacetime is described by the line element
\cite{Noteunits}
\begin{equation}\label{Eq1}
ds^2=-h(r)dt^2+{1\over{h(r)}}dr^2+r^2(d\theta^2+\sin^2\theta
d\phi^2)\  ,
\end{equation}
where the metric function $h(r)$ of a spherically symmetric black
hole of mass $M$ and electric charge $Q$ is given by
\begin{equation}\label{Eq2}
h(r)=1-{{2M}\over{r}}+{{Q^2}\over{r^2}}\  .
\end{equation}
The functional relation $h(r)=0$ determines the radii
$r_{\pm}=M\pm\sqrt{M^2-Q^2}$ of the black-hole (outer and inner)
horizons.

The composed curved-black-hole-spacetime-massless-scalar-field
system is characterized by the action \cite{Herch,Herr}
\begin{equation}\label{Eq3}
S=\int
d^4x\sqrt{-g}\Big[R-2\nabla_{\alpha}\phi\nabla^{\alpha}\phi-f(\phi){\cal
I}\Big]\ ,
\end{equation}
where the source term ${\cal I}$ is given by
\begin{equation}\label{Eq4}
{\cal I}=F_{\mu\nu}F^{\mu\nu}\  .
\end{equation}
Here $F_{\mu\nu}$ is the electromagnetic Maxwell tensor. The
non-minimal coupling between the scalar field and the
electromagnetic field of the charged black-hole spacetime is
determined by the coupling function $f(\phi)$. In the linearized
(weak-field) regime, the coupling function is characterized by the
universal quadratic behavior \cite{Herch,Herr}
\begin{equation}\label{Eq5}
f(\phi)=1-\alpha\phi^2\  ,
\end{equation}
where $\alpha$ is the dimensionless coupling parameter of the
theory. We shall henceforth assume $\alpha>0$.

The action (\ref{Eq3}), when varied with respect to the scalar field,
yields the scalar equation \cite{Herch,Herr}
\begin{equation}\label{Eq6}
\nabla^\nu\nabla_{\nu}\phi={1\over4}f_{,\phi}{\cal I}\  .
\end{equation}
The differential equation (\ref{Eq6}) determines the spatial and
temporal behaviors of the non-minimally coupled linearized scalar
field configurations in the charged black-hole spacetime.
Substituting the characteristic line element (\ref{Eq1}) of the
curved Reissner-Nordstr\"om spacetime into the scalar equation
(\ref{Eq6}) and using the field decomposition \cite{Notelm}
\begin{equation}\label{Eq7}
\phi(t,r,\theta,\phi)=\int\sum_{lm}{{\psi_{lm}(r;\omega)}\over{r}}Y_{lm}(\theta)e^{im\phi}e^{-i\omega
t} d\omega\  ,
\end{equation}
one obtains the Schr\"odinger-like differential equation
\begin{equation}\label{Eq8}
{{d^2\psi}\over{dy^2}}-V\psi=0\
\end{equation}
for the radial part of the scalar wave function. The effective binding potential in the
radial equation (\ref{Eq8}) is given by the expression
\cite{Herch,Herr}
\begin{equation}\label{Eq9}
V(r)=\Big(1-{{2M}\over{r}}+{{Q^2}\over{r^2}}\Big)\Big[{{l(l+1)}\over{r^2}}+{{2M}\over{r^3}}-{{2Q^2}\over{r^4}}
-{{\alpha Q^2}\over{r^4}}\Big]\  ,
\end{equation}
and the tortoise coordinate $y$ is defined by the relation
\cite{Notemap}
\begin{equation}\label{Eq10}
{{dr}\over{dy}}=h(r)\  .
\end{equation}

The linearized bound-state configurations of the nonminimally
coupled massless scalar fields in the charged Reissner-Nordstr\"om
black-hole spacetime (\ref{Eq1}) are determined by the radial
equation (\ref{Eq8}). This Schr\"odinger-like differential equation
is supplemented by the boundary conditions of spatially regular
scalar field configurations at the outer black-hole horizon and an
asymptotically decaying radial behavior at spatial infinity
\cite{Herch,Herr}:
\begin{equation}\label{Eq11}
\psi(r=r_+)<\infty\ \ \ \ ; \ \ \ \ \psi(r\to\infty)\to0\ .
\end{equation}

\section{The resonant spectrum of the composed
Reissner-Nordstr\"om-black-hole-massless-scalar-field system: A WKB
analysis}

In the present section we shall use analytical techniques in order
to determine the discrete resonant spectrum
$\{\alpha^-_n\}_{n=0}^{n=\infty}$ of the dimensionless coupling
constants that characterize the spatially regular composed
black-hole-nonminimally-coupled-scalar-field configurations in the
weak-field regime.

The spatially regular bound-state field configurations, which
characterize the Schr\"odinger-like differential equation
(\ref{Eq8}) with the effective radial potential (\ref{Eq9}), can be
studied analytically using standard WKB techniques. In particular, a
standard second-order WKB treatment for the bound states of the
effective binding potential (\ref{Eq9}) yields the compact
quantization condition \cite{WKB1,WKB2,WKB3,Note34}
\begin{equation}\label{Eq12}
\int_{y_{\text{in}}}^{y_{\text{out}}}dy\sqrt{-V(y;\alpha)}=\big(n-{1\over4}\big)\pi\
\ \ \ ; \ \ \ \ n=1,2,3,...,
\end{equation}
where the resonant parameter $n$ is an integer and
$\{y_{\text{in}},y_{\text{out}}\}$ are the radial turning points
[with $V(y_{\text{in}})=V(y_{\text{out}})=0$] of the binding
potential (\ref{Eq9}). [It should be noted that the WKB resonance
condition (\ref{Eq12}) fails in the extremal black-hole limit
$Q/M\to1$, in which case the effective binding potential (\ref{Eq9})
is characterized by a {\it degenerate} turning point at the
black-hole horizon $r=r_+$. In the next section we shall derive an
explicit analytical solution for the $l=0$ resonance equation of
\cite{Herch,Herr} in the near-extremal limit].

The WKB integral relation (\ref{Eq12}) determines the discrete
resonant spectrum $\{\alpha^-_n\}_{n=0}^{n=\infty}$ of the
dimensionless physical parameter $\alpha$ which characterizes the
spatially regular bound-state cloudy configurations of the
nonminimally coupled linearized scalar fields in the charged
black-hole spacetime. Taking cognizance of Eq. (\ref{Eq10}), one can
express the WKB resonance relation in terms of the radial coordinate
$r$:
\begin{equation}\label{Eq13}
\int_{r_{\text{in}}}^{r_{\text{out}}}dr{{\sqrt{-V(r;\alpha)}}\over{h(r)}}=\big(n-{1\over4}\big)\pi\
\ \ \ ; \ \ \ \ n=1,2,3,...\  .
\end{equation}
The radial turning points $\{r_{\text{in}},r_{\text{out}}\}$ of the
charged-black-hole-scalar-field binding potential $V(r;\alpha)$ are
determined by the relations [see Eq. (\ref{Eq9})]
\begin{equation}\label{Eq14}
1-{{2M}\over{r_{\text{in}}}}+{{Q^2}\over{r^2_{\text{in}}}}=0\
\end{equation}
and
\begin{equation}\label{Eq15}
{{l(l+1)}\over{r^2_{\text{out}}}}+{{2M}\over{r^3_{\text{out}}}}-{{2Q^2}\over{r^4_{\text{out}}}}
-{{\alpha Q^2}\over{r^4_{\text{out}}}}=0\  .
\end{equation}
From Eqs. (\ref{Eq2}), (\ref{Eq9}), (\ref{Eq14}), and (\ref{Eq15}), one
deduces that, in the eikonal large-coupling (large-$\alpha$) regime,
the WKB integral relation (\ref{Eq13}) can be approximated by
\begin{equation}\label{Eq16}
\int_{r_+}^{\infty}dr{{\sqrt{{{\alpha
Q^2}\over{r^4\Big(1-{{2M}\over{r}}+{{Q^2}\over{r^2}}\Big)}}}}}=\big(n+{3\over4}\big)\pi\
\ \ \ ; \ \ \ \ n=0,1,2,...\ .
\end{equation}
Interestingly, the integral on the l.h.s of the WKB resonance
equation (\ref{Eq16}) can be evaluated analytically to yield the
remarkably compact resonant spectrum
\begin{equation}\label{Eq17}
\alpha^-_n=\Bigg[{{2\pi}\over{\ln{\Big({{1+{\bar Q}}\over{1-{\bar
Q}}}\Big)}}}\Bigg]^2\cdot \big(n+{3\over4}\big)^2\ \ \ \ ; \ \ \ \
n=0,1,2,...\
\end{equation}
which characterizes the composed bound-state
charged-black-hole-nonminimally-coupled-massless-scalar-field
configurations along the critical existence line of the system.

It is worth emphasizing the fact that, formally, the discrete WKB
resonant spectrum  (\ref{Eq17}) is expected to be valid in the
eikonal large-$n$ (large-$\alpha$) regime. However, below we shall
explicitly show that, in the dimensionless physical regime ${\bar
Q}\lesssim0.8$ of the central supporting black hole, the
approximated ({\it analytically} derived) resonant spectrum
(\ref{Eq17}) agrees remarkably well with the corresponding exact
({\it numerically} computed) resonant spectrum \cite{Herch,Herr} even
in the regime $n=O(1)$ of the fundamental resonant modes.

\section{Non-minimally coupled scalar field configurations supported by near-extremal
charged black holes: Analytic treatment}

As shown in \cite{Herch,Herr}, for spherically-symmetric cloudy
scalar field configurations, the existence line of the composed
black-hole-scalar-field system is determined by the resonance
equation
\begin{equation}\label{Eq18}
{_2F_1}\big[1/2-\sqrt{1-4\alpha}/2/2,1/2+\sqrt{1-4\alpha};1;x^2/(x^2-1)\big]=0\
,
\end{equation}
where $x\equiv Q/r_+$. This equation has been solved numerically in
\cite{Herch,Herr}. In the present section we shall explicitly show
that the resonance equation (\ref{Eq18}) can be solved {\it
analytically} in the regime
\begin{equation}\label{Eq19}
\epsilon\equiv 1-{\bar Q}\ll1\
\end{equation}
of near-extremal supporting black-hole spacetimes.

We first point out that, using equation 15.3.7 of \cite{Abram}, one
can express the resonance equation (\ref{Eq18}) in the form
\begin{eqnarray}\label{Eq20}
{{\Gamma(i\beta)}\over{[\Gamma(1/2+i\beta/2)]^2}}\cdot\big(\sqrt{8\epsilon}\big)^{1/2-i\beta/2}
%\nonumber \\&& \times
\cdot{_2F_1}\big(1/2-i\beta/2,1/2-i\beta/2;1-i\beta;-\sqrt{8\epsilon}\big)+(\beta\to
-\beta)=0\ ,
\end{eqnarray}
where
\begin{equation}\label{Eq21}
\beta\equiv -i\sqrt{1-4\alpha}\in\mathbb{R}\ ,
\end{equation}
and the notation $(\beta\to -\beta)$ means ``replace $\beta$ by
$-\beta$ in the preceding term". Taking cognizance of the
near-extremal relation (\ref{Eq19}) and using the simple asymptotic
behavior (see Eq. 15.1.1 of \cite{Abram})
\begin{equation}\label{Eq22}
_2F_1(a,b;c;z)\to 1\ \ \ \text{for}\ \ \ {{ab}\over{c}}\cdot z\to 0\
\end{equation}
of the hypergeometric function, one can approximate the resonance
equation (\ref{Eq20}) in the near-extremal $\epsilon\ll1$ regime by
the mathematical relation
\begin{eqnarray}\label{Eq23}
{{\Gamma(i\beta)}\over{[\Gamma(1/2+i\beta/2)]^2}}\cdot\big(\sqrt{8\epsilon}\big)^{1/2-i\beta/2}
%\nonumber \\&& \times
+{{\Gamma(-i\beta)}\over{[\Gamma(1/2-i\beta/2)]^2}}\cdot\big(\sqrt{8\epsilon}\big)^{1/2+i\beta/2}=0\  ,
\end{eqnarray}
which yields
\begin{equation}\label{Eq24}
\big(\sqrt{8\epsilon}\big)^{-i\beta}=-{{\Gamma(-i\beta)[\Gamma(1/2+i\beta/2)]^2}\over{\Gamma(i\beta)[\Gamma(1/2-i\beta/2)]^2}}\
.
\end{equation}
Assuming $\beta\ll1$ in the near-extremal regime (\ref{Eq19}) [see Eq. (\ref{Eq26}) below], one can
approximate the (rather cumbersome) resonance equation (\ref{Eq24}) by
the simpler functional relation \cite{Noteher}
\begin{equation}\label{Eq25}
(8\epsilon)^{-i\beta/2}=1-i4\ln2\cdot\beta+O(\beta^2)\  .
\end{equation}

The mathematical relation (\ref{Eq25}) determines the functional dependence of the
dimensionless physical parameter $\beta$ on the dimensionless
charge-to-mass ratio of the central supporting black hole. In
particular, from (\ref{Eq25}) one finds the relation \cite{Noteintg}
\begin{equation}\label{Eq26}
\beta^+_n={{4\pi n}\over{\ln\Big({{32}\over{1-{\bar Q}}}\Big)}} \ \
\ ; \ \ \ n=1,2,3,...\ \ \ \ \text{for}\ \ \ \ \beta\ll1\  .
\end{equation}
Taking cognizance of Eq. (\ref{Eq21}), one obtains from (\ref{Eq26}) the
compact functional expression
\begin{equation}\label{Eq27}
\alpha^-_n={1\over4}+\Bigg[{{2\pi n}\over{\ln\Big({{1-{\bar
Q}}\over{32}}\Big)}}\Bigg]^2\ \ \ ; \ \ \ n=1,2,3,...\ \ \ \
\text{for}\ \ \ \ 1-{\bar Q}\ll1\
\end{equation}
for the discrete resonant spectrum which characterizes the composed
near-extremal-black-hole-linearized-massless-scalar-field
configurations.

\section{Numerical confirmation}

In the present section we shall test the accuracy of the
analytically derived resonant spectra (\ref{Eq17}) and (\ref{Eq27}),
which characterize the composed
Reissner-Nordstr\"om-black-hole-nonminimally-coupled-massless-scalar-field
system, against the corresponding exact resonant spectra as computed
numerically from equation (\ref{Eq18}) (this resonant equation was
first derived in \cite{Herch,Herr}).

In Table \ref{Table1} we display, for various values of the
dimensionless charge-to-mass ratio of the central supporting black
hole, the dimensionless ratio ${\cal R}_n({\bar Q})\equiv
\alpha^{\text{analytical}}_n/\alpha^{\text{numerical}}_n$ between
the approximated ({\it analytically} calculated) discrete values
$\{\alpha^-_n({\bar Q})\}_{n=0}^{n=\infty}$ of the dimensionless
coupling parameter $\alpha$, which characterizes the composed
black-hole-scalar-field system, and the corresponding exact ({\it
numerically} computed) values of the coupling parameter. From the
data presented in Table \ref{Table1} one learns that the agreement
between the analytically derived WKB resonant spectrum (\ref{Eq17})
and the corresponding numerically computed coupling parameters
\cite{Herch,Herr} is remarkably good in the eikonal large-$\alpha$
(large-$n$) regime of the discrete resonant spectrum. In fact, the
agreement between the analytical and numerical results is found to
be quite good already in the fundamental $n=O(1)$ regime.

\begin{table}[htbp]
\centering
\begin{tabular}{|c|c|c|c|c|c|c|c|}
\hline \ \ $Q/M$\ \ & \ $n=0$\ \ & \ $n=1$\ \ & \
$n=2$\ \ & \ $n=3$\ \ & \ $n=4$\ \ & \ $n=5$\ \ & \ $n=6$\ \ \\
\hline \ \ $0.2$\ \ \ &\ \ $0.958$\ \ \ &\ \ $0.992$\ \ \ &\ \
$0.996$\ \ \ &\ \ $0.998$\ \ \ &\ \ $0.999$\ \ \ &\ \ $0.999$\ \ \ &\ \ $0.999$\ \ \\
\hline \ \ $0.5$\ \ \ &\ \ $0.944$\ \ \ &\ \ $0.989$\ \ \ &\ \
$0.995$\ \ \ &\ \ $0.997$\ \ \ &\ \ $0.998$\ \ \ &\ \ $0.999$\ \ \ &\ \ $0.999$\ \ \\
\hline \ \ $0.8$\ \ \ &\ \ $0.898$\ \ \ &\ \ $0.981$\ \ \ &\ \
$0.992$\ \ \ &\ \ $0.995$\ \ \ &\ \ $0.997$\ \ \ &\ \ $0.998$\ \ \ &\ \ $0.999$\ \ \\
\hline
\end{tabular}
\caption{The discrete resonant spectrum of the composed
charged-black-hole-nonminimally-coupled-linearized-massless-scalar-field
configurations. We present the charge-dependent dimensionless ratio
${\cal R}_n({\bar Q})\equiv
\alpha^{\text{analytical}}_n/\alpha^{\text{numerical}}_n$ between
the analytically derived WKB resonant spectrum (\ref{Eq17}), which
characterizes the dimensionless coupling parameter $\alpha$ of the
theory, and the corresponding exact values of the coupling parameter
as computed numerically from the $l=0$ resonance equation (\ref{Eq18}) \cite{Herch,Herr}. One
finds the characteristic relation ${\cal R}_n({\bar Q})\to1$ in the
eikonal large-$\alpha$ (large-$n$) regime of the resonant spectrum.
Interestingly, one finds that in the physical regime ${\bar Q}\lesssim 0.8$ of the central supporting black hole, the
agreement between the analytical and numerical results is quite good already in the $n=O(1)$ regime.}
\label{Table1}
\end{table}

In Table \ref{Table2} we present the fundamental coupling constants
$\alpha^{\text{analytical}}_0(\epsilon)$, which characterize the
composed black-hole-nonminimally-coupled-scalar-field system in the
regime $\epsilon\equiv 1-{\bar Q}\ll1$ of near-extremal charged
supporting black holes, as calculated from the analytically derived
resonant spectrum (\ref{Eq27}). We also present the exact
(numerically computed) coupling constants
$\alpha^{\text{numerical}}_0(\epsilon)$ of the composed system as
computed numerically from the $l=0$ resonant equation (\ref{Eq18})
\cite{Herch,Herr}. The data presented in Table \ref{Table2} reveals
the fact that the agreement between the analytically calculated and
numerically computed resonant spectra becomes extremely good in the
extremal $\epsilon\to0$ limit.

\begin{table}[htbp]
\centering
\begin{tabular}{|c|c|c|c|c|c|c|c|}
\hline \ \ $\epsilon\equiv 1-Q/M$\ \ & \ $10^{-4}$\ \ & \ $10^{-5}$\
\ & \ $10^{-6}$\ \ & \ $10^{-7}$\ \ & \ $10^{-8}$\
\ & \ $10^{-9}$\ \ & \ $10^{-10}$\ \ \\
\hline \ \ $\alpha^{\text{analytical}}_0(\epsilon)$\ \ &\ \
$0.496$\ \ \ &\ \ $0.426$\ \ \ &\ \ $0.382$\ \ \ &\ \ $0.353$\ \ \ &\ \ $0.332$\ \ \ &\ \ $0.317$\ \ \ &\ \ $0.306$\ \ \\
\hline \ \ $\alpha^{\text{numerical}}_0(\epsilon)$\ \ &\ \
$0.530$\ \ \ &\ \ $0.442$\ \ \ &\ \ $0.390$\ \ \ &\ \ $0.357$\ \ \ &\ \ $0.335$\ \ \ &\ \ $0.318$\ \ \ &\ \ $0.307$\ \ \\
\hline
\end{tabular}
\caption{Nonminimally coupled scalar field configurations supported by near-extremal
charged black holes. We present the fundamental coupling constants
$\alpha^{\text{analytical}}_0(\epsilon)$, which characterize the composed
black-hole-massless-scalar-field system in the near-extremal regime
$\epsilon\equiv 1-{\bar Q}\ll1$ of the central supporting black holes,
as obtained directly from the analytically derived resonant spectrum
(\ref{Eq27}). Also presented are the corresponding exact coupling
constants $\alpha^{\text{numerical}}_0(\epsilon)$ of the theory as
computed numerically from the resonance equation (\ref{Eq18})
\cite{Herch,Herr}. One finds a remarkably good agreement between the
analytically calculated and numerically computed resonant spectra in
the near-extremal $\epsilon\ll1$ regime.} \label{Table2}
\end{table}

\section{Summary}

The recently published important works \cite{Herch,Herr} have
revealed the interesting fact that charged black holes can support
spatially regular hairy configurations made of massless scalar
fields which are characterized by nontrivial (nonminimal) couplings
[see Eq. (\ref{Eq3})] to the electromagnetic fields of the central
supporting black holes. Intriguingly, it has been demonstrated
numerically \cite{Herch,Herr} that this highly interesting physical
phenomenon, known by the name black-hole spontaneous scalarization,
can only occur when the dimensionless coupling parameter $\alpha$
\cite{Notealpha} of the theory belongs to a discrete
charge-dependent resonant spectrum $\alpha\in\{[\alpha^{-}_{n}({\bar
Q}),\alpha^{+}_{n}({\bar Q})]\}_{n=0}^{n=\infty}$ of scalarization
bands. In particular, the discrete set $\{\alpha^{-}_{n}({\bar
Q})\}_{n=0}^{n=\infty}$ of resonant modes, which characterizes the
physical coupling parameter $\alpha$, corresponds to the regime of
spatially regular linearized scalar field configurations (scalar
clouds) which are supported by the central charged black holes.

In the present paper we have explicitly shown that the
physical and mathematical properties of the spatially regular cloudy scalar field
configurations, which are non-minimally coupled to the electromagnetic field of
the central supporting charged black hole, can be studied analytically.
In particular, using a WKB analysis, we have derived the remarkably
compact discrete resonant spectrum (\ref{Eq17}) for the dimensionless physical parameter
$\alpha$ which characterizes the composed
charged-black-hole-linearized-massless-scalar-field configurations.
Furthermore, we have explicitly demonstrated that, for
spherically symmetric field configurations, the {\it analytically}
derived resonant spectrum (\ref{Eq17}) agrees remarkably well (see the
data presented in Table \ref{Table1}) with the exact resonant
spectrum as computed {\it numerically} from the $l=0$ resonance
equation (\ref{Eq18}) \cite{Herch,Herr}.

Finally, we would like to note that the discrete resonant spectrum
(\ref{Eq17}) yields the remarkably simple analytical expression \cite{Note00,Notene}
\begin{equation}\label{Eq28}
{\bar Q}_{\text{min}}(\alpha)={{1-{\cal F}(\alpha)}\over{1+{\cal
F}(\alpha)}}\ \ \ \ \text{with}\ \ \ \ {\cal F}(\alpha)\equiv
e^{-3\pi/2\sqrt{\alpha}}\
\end{equation}
for the {\it existence-line} which characterizes the composed
charged-black-hole-nonminimally-coupled-scalar-field system. It is
worth emphasizing again that the physical significance
\cite{Herch,Herr} of the $\alpha$-dependent existence line
(\ref{Eq28}) stems from the fact that this critical line marks the
boundary between spontaneously scalarized charged hairy black-hole
configurations and bald Reissner-Nordstr\"om black holes
\cite{Noteael}.

\bigskip
\noindent
{\bf ACKNOWLEDGMENTS}
\bigskip

This research is supported by the Carmel Science Foundation. I would
like to thank Yael Oren, Arbel M. Ongo, Ayelet B. Lata, and Alona B.
Tea for helpful discussions.

%\newpage


\begin{thebibliography}{99}
\bibitem{NoteMQ} Here $\{M,Q\}$ are respectively the mass and electric charge of the charged black-hole spacetime.
We shall henceforth assume $Q>0$ without loss of generality.
\bibitem{Noteunits} We shall use natural units in which $G=c=1$.
\bibitem{Notelm} The integer indices $\{l,m\}$ are the dimensionless angular harmonic
parameters which characterize the non-minimally coupled linearized
scalar field modes in the charged Reissner-Nordstr\"om black-hole
spacetime. For brevity, we shall henceforth omit these harmonic
field parameters.
\bibitem{Notemap} Note that the radial regime $r\in[r_+,\infty]$ is
mapped by the differential relation (\ref{Eq10}) to the regime
$y\in[-\infty,\infty]$.
\bibitem{Noteher} Here we have used equations 6.1.8, 6.3.1, and 6.3.3 of \cite{Abram}.
\bibitem{Noteintg} Here we have used the relation $1=e^{2i\pi n}$,
where the resonance parameter $n$ is an integer.
\bibitem{Notene} As emphasized above, the WKB analysis is expected
to break down in the regime $\alpha_{\text{min}}=O(1)$ of
near-extremal black holes. In this limiting case we have explicitly
shown that the $l=0$ resonance condition (\ref{Eq18}) of
\cite{Herch,Herr} is amenable to an analytical treatment. In
particular, from the analytically derived resonant spectrum
(\ref{Eq27}) one obtains the existence line ${\bar
Q}_{\text{min}}(\alpha)=1-32e^{-2\pi/\sqrt{\alpha-1/4}}$ in the
near-extremal regime of the central supporting black holes.
\bibitem{Noteael} Note, in particular, that for a given value of the dimensionless
coupling parameter $\alpha$, the charged hairy black-hole spacetimes
are characterized by the relation ${\bar Q}(\alpha)\geq {\bar Q}_{\text{min}}(\alpha)$.
\end{thebibliography}
\end{document}